\documentstyle[amssymb,11pt]{article}
\normalbaselines \oddsidemargin 0.5cm \evensidemargin 0.5cm
\begin{document}

\title{Unruh's detector in the presence of Lorentz symmetry breaking}
\author{R. Rashidi, N. Khosravi, E. Khajeh and H. Salehi\\
%EndAName
{ {Department of Physics, Shahid Beheshti University, Evin,
Tehran 19839, IRAN}}\\\\
{E-mails: r-rashidi, n-khosravi, e-khajeh, h-salehi@sbu.ac.ir}}
\date{}
\maketitle

\begin{abstract}
We investigate the quantum field theory of a Lorentz non-invariant
model with a massive nonlinear dispersion relation in Minkowski
space. The model involves some non-causal signals in the form of
wave packets propagating with super-luminal group velocities. To
avoid the problems with causality we characterize the causal sector
of the theory by a cutoff condition excluding all super-luminal
group velocities. It is argued that in the causal theory satisfying
the energy positivity condition an Unruh's detector moving with a
constant velocity with respect to the preferred frame does not
detect any particle. But in a causal theory violating energy
positivity the registered of field quanta occurs. We comments on the
origin of this particle creation.

\vspace{3mm}\noindent {\LARGE {\bf \ }}
\end{abstract}

\section{Introduction}

The Lorentz symmetry is supposed to be a fundamental symmetry of
physics. It contains two distinct sub-groups, rotation and boost.
The former has not any ambiguity since it can be checked and has
been checked. But boost transformations is not clear very well.
Since their parameters are unbounded, the Lorentz group is
non-compact. This non-compactness imposes an upper-bound for testing
the Lorentz symmetry experimentally. It means that the Lorentz
symmetry may be a symmetry in low energy physics only. This aspect
is clearly reflected in the theoretical apparat of Quantum Field
Theory where a short distance cutoff is always necessary to remove
the divergencies or even to avoid the triviality problem \cite{1}.
Such a cutoff may break the Lorentz symmetry\footnote{This argument
is not appropriate for any theory, for example for loop quantum
gravity, see \cite{2b}.}. Quantum Gravity and String Theory predict
also some breaking of the Lorentz symmetry \cite{2,3,9}. This arises
as a result of the existence of a minimum length \cite{4}. In
addition, some cosmological observations suggest some deviation from
conventional energy-momentum's dispersion relation of particles
predicted by spacial relativity e.g.: events involving gamma
radiation with energies beyond 20TeV from distant sources such as
Markarian 421 and Markarian 501 blazars \cite{6} and studies of the
evolution of air showers produced by ultra high-energy hadronic
particles suggest that pions live longer than expected \cite{7}.
These experimental data, if confirmed (observation uncertainty is
still too large), might indicate that Lorentz invariance is
violated\footnote{In this context one should also mention those
experiments including resonant cavity tests and clock-comparison
experiments in \cite{c3,c4,c5}. But the results do not indicate any
Lorentz violating effects and merely improve the constraints on the
parameters of the Lorentz violating standard model extension
\cite{3a}.}. This, however, may be one possibility among others
\cite{c1,c2}, but taking it as a serious one, so one has to deal
with the modification of the dispersion relations due to Lorentz
violation \cite{8}.

In this paper we study the Lorentz symmetry breaking in the context
of Quantum Field Theory in a classical background. Specifically we
study the response of an Unruh's detector in Minkowski space. If
such a detector moves with a constant acceleration with respect to
some inertial system, it detects a thermal radiation (Unruh's
effect) \cite{unruh}. The same detector however does not detect any
particle if it moves with a constant velocity. This arises as a
consequence of the exact Lorentz invariance of the vacuum of
Minkowski space. One may ask how this latter result is changed in
the presence of the Lorentz symmetry breaking? In the remainder of
this paper we will try to answer this question. In section 2 we
describe a model of Lorentz symmetry breaking. Unruh's detector will
be discussed in section 3 and the effect of symmetry breaking will
be demonstrated. In the last section we will close this paper with
some concluding remarks.

\section{The Lorentz symmetry breaking model}

We begin with the massive version of Jacobson and et al.'s model for
breaking the boost sub-group of the Lorentz group \cite{11}. In this
model, the Lorentz invariance is broken using a dynamical coupling
of a real scalar field $\varphi $ with a preferred frame. The
Lagrangian of this model is given by
\begin{equation}
{\cal L}_{\varphi }=1/2(\nabla ^{a}\varphi \nabla _{a}\varphi
+\alpha ^{2}(D^{2}\varphi )^{2}-m^2 \varphi^2)\hspace{0.15cm}
\label{b}
\end{equation}%
where $\alpha $ is a real constant with a very small value and
$D^{2}$ is defined by
\begin{equation}
D^{2}\varphi =-D^{a}D_{a}\varphi =-q^{ac}\nabla _{c}(q_{a}^{b}\nabla
_{b}\varphi )\hspace{0.15cm}  \label{c}
\end{equation}%
with
\begin{equation}
q_{ab}=-\eta_{ab}+u_{a}u_{b}\hspace{0.15cm}  \label{d}
\end{equation}%
where $\eta_{ab}=diag(1,-1,-1,-1)$ is the four dimensional
Lorentzian metric and the vector $u_{a}$ is a unit future directed
time-like vector field. It can be interpreted as the four velocity
of a preferred inertial observer. The rest frame of this observer
may be considered as the mathematical realization of a preferred
frame (aether). Here we assume this preferred frame be the rest
frame of a preferred inertial observer thus we set $u_{a}=(1,0,0,0)$
in the preferred frame.

We distinguish between two different kinds of transformations
according to the work \cite{3a} and \cite{ko}, namely the observer
Lorentz transformation and the particle Lorentz transformation. The
former corresponds to a transformation of an inertial system while
the latter corresponds to a transformation of particles or localized
fields within a fixed inertial system.

 Since the Lagrangian (\ref{b}) is a space-time scalar under observer transformations, the theory exhibits observer Lorentz symmetry.
 But this Lagrangian is not invariant under the
particle boost transformation
\begin{equation}\label{m1}
    \varphi'(x)=\varphi(\Lambda^{-1}x)
\end{equation}
where $\Lambda$ is the boost transformation matrix . Hence the
Lagrangian (\ref{b}) violates the particle Lorentz invariance. This
is taken as acceptable form of Lorentz violation for physical
systems\footnote{Particle boost transformation are sometimes called
active Lorentz transformation \cite{da}}. In the present context any
particle Lorentz transformation may be imagined as a transformation
of a physical system (particles or localized field) within the rest
frame of the preferred observer, i.e. the aether.

We note that the symmetry under space-time translations, considered
here as particle (active) transformations
\begin{equation}\label{m2}
    \varphi'(x)=\varphi(x-a),
\end{equation}
where $a$ is a real four-vector, is valid.

 In the preferred frame the equation of motion takes the from
\begin{equation}\label{g}
    \Box\varphi-\alpha^2q^{ab}q^{cd}\nabla_a\nabla_b\nabla_c\nabla_d\varphi+m^2\varphi=0\hspace{.15cm}.
\end{equation}
The positive frequency solution of (\ref{g}) are
\begin{equation}\label{h}
    u_{k}(x)\propto\exp(-ik_{a}x^{a})\hspace{.15cm}
\end{equation}
where $k^{a}$ and $x^{a}$ are 4-vectors and $k_{a}$ satisfy the
modified dispersion relation
\begin{equation}\label{i}
    \eta^{ab}k_ak_b-\alpha^2q^{ab}q^{cd}k_ak_bk_ck_d-m^2=\omega^{2}-k^{2}+\alpha^{2}k^{4}-m^2=0\hspace{.15cm}.
\end{equation}
The 4-vector solutions of this dispersion relation at sufficiently
high frequencies become space-like (at the massless limit all
solutions are space-like). These solutions violate the energy
positivity condition ($\omega\geq|k|$) and can introduce some
instabilities in the quantized theory \cite{4a}.
 Moreover, this nonlinear dispersion relation means that the group
velocity of the wave, namely,
\begin{equation}\label{i2}
v_g=\frac{d\omega}{d|\vec{k}|}=\frac{1-2\alpha^2k^2}{\sqrt{1-\alpha^2k^2+m^2/k^2}},
\end{equation}
is not a constant and, at sufficiently high frequencies, can be
greater than the velocity of light ($c=1$). This leads to non-local
wave packets which violate causality. Since the quantization of such
wave packets is obscure one can define a causal sector of the theory
by the condition that all super-luminal group velocities are
excluded. This naturally imposes a cutoff $k_c$ on the momentum
space and ensures us that all wave packets propagate with a
sub-luminal (or luminal) velocity. In this paper, we assume $\alpha
m\ll1$ and take the cutoff bound as $k_c=\sqrt{\frac{m}{\alpha}}$.
One can see that for all $|\vec{k}|$ below $k_c$ the wave-vector
$k^\mu$ is time-like and the group velocity does not exceed $1$.Thus
this causal sector satisfies the energy positivity condition.
Furthermore, If
$k^2>\frac{1+\sqrt{1+4\alpha^2m^2}}{2\alpha^2}>\frac{m}{\alpha}$ the
frequency $\omega$ will take an imaginary value. It causes the mode
solutions become singular. When imaginary frequency exist, the
energy spectrum of the field theory is unbounded from below, a
condition which is generally believed to lead to instability
\cite{ff}. Moreover, the existence of such imaginary frequencies can
cause some problems concerning the unitarity of the time-evolution
operator. But this region of the momentum space is above the chosen
cutoff and therefore these imaginary frequency modes are
automatically eliminated.

The scalar product for two solutions of equation of motion (\ref{g}) is
defined as
\begin{equation}
\begin{array}{l}
\vspace{0.2cm}(\varphi _{1},\varphi _{2})=-i\int_{\Sigma }d\Sigma _{a}%
\hspace{0.2cm}\{\varphi _{1}(x)(g^{ab}\nabla \hspace{-0.3cm}%
^{^{^{\leftrightarrow }}}\hspace{0cm}_{b}-\alpha ^{2}q^{ab}q^{cd}\hspace{%
0.1cm}{\nabla _{b}\nabla _{c}\nabla _{d}}\hspace{-1.3cm}^{^{^{\longleftarrow
}}}\hspace{-0.11cm}^{^{^{-}}}\hspace{-0.11cm}^{^{^{\longrightarrow }}}%
\hspace{0.5cm})\varphi _{2}^{\ast }(x)\} \\
\hspace{1.42cm}:=-i\int_{\Sigma }d\Sigma _{a}\hspace{0.2cm}{\cal J}^{a}%
\hspace{0.15cm}%
\end{array}
\label{j}
\end{equation}%
where
\begin{equation}
\begin{array}{l}
\varphi _{1}{\nabla _{a}\nabla _{b}\nabla _{c}}\hspace{-1.3cm}%
^{^{^{\longleftarrow }}}\hspace{-0.11cm}^{^{^{-}}}\hspace{-0.11cm}%
^{^{^{\longrightarrow }}}\hspace{0.5cm}\varphi _{2}= \\
+\varphi _{1}{\nabla _{a}\nabla _{b}\nabla _{c}}\varphi _{2}-\varphi _{2}{%
\nabla _{a}\nabla _{b}\nabla _{c}}\varphi _{1} \\
-\nabla _{a}\varphi _{1}\nabla _{b}\nabla _{c}\varphi _{2}+\nabla
_{a}\varphi _{2}\nabla _{b}\nabla _{c}\varphi _{1} \\
+\nabla _{b}\varphi _{1}\nabla _{a}\nabla _{c}\varphi _{2}-\nabla
_{b}\varphi _{2}\nabla _{a}\nabla _{c}\varphi _{1} \\
-\nabla _{c}\varphi _{1}\nabla _{b}\nabla _{a}\varphi _{2}+\nabla
_{c}\varphi _{2}\nabla _{b}\nabla _{a}\varphi _{1}\hspace{0.15cm}.%
\end{array}
\label{n}
\end{equation}%
With this definition we have $\nabla _{a}{\cal {J}}^{a}=0$ \footnote{%
Note that the complex form of the Lagrangian (\ref{b}) is invariant under
global $U(1)$ transformation obviously. So as a consequence of Noether's
theorem a conserved current $j^{a}$ exists. The form of this current $j^{a}$
corresponds to the form of current ${\cal {J}}^{a}$ introduced in (\ref{j}).}%
. As a consequence the scalar product (\ref{j}) becomes time independent.
With respect to (\ref{j}) the orthonormal positive frequency mode solutions
of (\ref{g}) are
\begin{equation}
u_{\vec{k}}(x)=\frac{1}{\sqrt{2k_{0}-4\alpha
^{2}q_{0b}k^{b}(q_{cd}k^{c}k^{d})}}\exp (-ikx)\hspace{0.15cm}.  \label{o}
\end{equation}%
In the casual sector the field $\varphi(x)$ is taken to expanded in
terms of $u_{k}(x)$ and $u^*_{k}(x)$ with $|k|$ not exceeding the
cutoff $k_c$, namely
\begin{equation}
\varphi (x)=\int_{|\vec{k}|\leq k_c} (a_{\vec{k}}u_{\vec{k}}(x)+a_{\vec{k}}^{\dag }u_{\vec{k}%
}^{\ast }(x))d^{3}k.\label{p}
\end{equation}%

\section{Quantization}

 To quantize the field
represented by (\ref{p}) we begin with the following general
remarks. Since the particle boost transformations (\ref{m1}) are not
symmetry transformations of the Lagrangian (\ref{b}), at the quantum
level these transformations can not unitarily be implemented in a
given representation. However as mentioned above, the Lagrangian
(\ref{b}) is invariant under the space-time translations (\ref{m2}).
Thus space-time translations can unitarily be implemented. Because
these symmetry transformations leave the vector field $u_{a}$
invariant, one can consider $u_{a}$ as a super-selection quantity
that takes a particular value, namely $u_{a}=(1,0,0,0)$ within a
corresponding unitary class of a given representation. This unitary
representation will be called the preferred representation of the
model.

The unitary class of the preferred representation can be obtained in
terms of the Fock space representation as follows. According to the
invariance of the Lagrangian (\ref{b}) under the space-time
translations (\ref{m2}) one can obtain a canonical energy-momentum
tensor which is conserved for solutions of the equation of motion:

\begin{equation}\label{a1}
T^{\mu\nu}= \frac{\partial{\cal
L}}{\partial(\partial_{\mu}\varphi)}\partial^{\nu}\varphi+\frac{\partial\cal
L}{\partial(\partial_{\mu}\partial_{\rho}\varphi)}\partial_{\rho}\partial^{\nu}\varphi-\partial_{\rho}(\frac{\partial\cal
L}{\partial(\partial_{\mu}\partial_{\rho}\varphi)})\partial^{\nu}\varphi-{\cal
L }\eta^{\mu\nu}\hspace{.2cm} (\mu,\nu =0,..,3)
\end{equation}
and
\begin{equation}\label{a2}
\partial_{\mu}T^{\mu\nu}=0
\end{equation}
for all solutions of equation (\ref{g}). The Integral of $T^{0\mu}$
over the space-like hypersurface $t=const.$ represents the
time-conserved energy-momentum four-vector
\begin{equation}\label{a3}
P^{\mu}=\int T^{0\mu}d^3x.
\end{equation}

To quantized this model, we note that the space-time translations
(\ref{m2}), as the symmetry of the model, can unitarily be
implemented. This means that $P^{\mu}$ can be considered as the
generators of space-time translations, yielding
\begin{equation}\label{a4}
[P^{\mu}, \varphi(x)]=i \partial^{\mu}\varphi(x)
\end{equation}
which determines the operator nature of $\varphi(x)$. In particular,
the equations (\ref{a4}) guarantee that time translation is a
unitary transformation.

 We note that the quantization rule based
on (\ref{a4}) is an application of the correspondence principle and
is more general than the standard canonical quantization, because it
uses only the symmetry transformation of the theory\footnote{For a
linear theory without a (momentum) cutoff (\ref{a4}) is equivalent
with
 the standard form of the canonical method \cite{bog}.}.
Therefore, it is applicable to all kind of theories admitting the
space-time translation invariance such as the theory considered in
this paper. In the present context, the application of the
quantization rule (\ref{a4}) is urgent, because a priori it is not
known whether the standard canonical quantization scheme is
appropriate in the presence of the cutoff.

 Now taking (\ref{p}) and (\ref{a4}) together we get
\cite{bog}
\begin{equation}\label{r}
    [a_{\vec{k}},a_{\vec{k'}}]=[a^\dag_{\vec{k}},a^\dag_{\vec{k'}}]=0 \hspace{.15cm} ,\hspace{.15cm}
    [a_{\vec{k}},a^\dag_{\vec{k'}}]=\delta^3(\vec{k}-\vec{k'})\hspace{.15cm}
\end{equation}
which are the usual form of commutation relations for $a_{\vec{k}}$
and $a^\dag_{\vec{k}}$. Thus we can consider $a_{\vec{k}}$ and
$a^\dag_{\vec{k}}$, as annihilation and creation operators of a
particle with momentum $\vec{k}$ and energy $\omega$, respectively.
One can define a no particle state as
\begin{equation}\label{s}
    a_{\vec{k}}\mid0>=0\hspace{.15cm},\forall \vec{k} \hspace{.10cm} (|\vec{k}|\leq k_c)
\end{equation}
and a one-particle state as
\begin{equation}\label{s1}
\mid \vec{k}>=a^{\dag}_{\vec{k}}\mid0>
\end{equation}
which is a eigenstate of energy-momentum operator:
\begin{equation}\label{s2}
P^{\mu}\mid \vec{k}>=k^{\mu}\mid \vec{k}>.
\end{equation}
A general n-particle state may be constructed in the standard
manner.

\section{Unruh's detector}

We consider now an Unruh's detector \cite{12}, understood here as a
point-like detector moving along a general trajectory $x=x(\tau )$
interacting with the scalar field (\ref{p}). The interaction between
detector and the field is as follows:
\begin{equation}
\delta {\cal H}=\beta \hat{M}(\tau )\varphi (x(\tau
))\hspace{0.15cm} \label{t}
\end{equation}%
where $\beta $ is coupling constant, $\tau $ is detector proper time and $%
\hat{M}(\tau )$ is a quantum operator that describes the detector's
monopole moment. Let the detector ground state be $\left\vert
E_{0}\right\rangle $ and its possible states form $\left\vert
{E}\right\rangle $. In this sense, the absorption of a particle
changes the state of the detector from $\left\vert
E_{0}\right\rangle$ to $\left\vert E\right\rangle$ (such that
$\Delta E=E-E_{0}>0$) and $\Delta E<0$ for emission of a particle
with respect to the detector. Transition amplitude ${\cal M}$ from
an initial state $\left\vert E_{i}\right\rangle \left\vert
i\right\rangle $ to a final state $\left\vert E_{f}\right\rangle
\left\vert f\right\rangle $ at the first order perturbation is
($\left\vert i\right\rangle $ and $\left\vert f\right\rangle $ are
scalar field states)
\begin{equation}
{\cal M}(i,f)={\cal M}(E_{i},\tau _{i};E_{f},\tau _{f})=\beta
E_{i,f}\int_{\tau _{i}}^{\tau _{f}}d\tau ^{\prime}e^{
-i(E_{f}-E_{i})\tau^{\prime }}\left\langle f\right\vert \varphi
(x(\tau^{\prime }))\left\vert i\right\rangle \hspace{0.15cm}  \label{u}
\end{equation}%
where $E_{i,f}=\left\langle E_{f}\right\vert \hat{M}(0)\left\vert
E_{i}\right\rangle $ ($\hat{M}(\tau )=e^{i\hat{{\cal H}}_{0}\tau }\hat{M}%
(0)e^{-i\hat{{\cal H}}_{0}\tau }$ and $\hat{{\cal H}}_{0}$ is the free
Hamiltonian of the detector). Then probability of this transition is
\begin{equation}
{\cal P}_{i,f}=\beta ^{2}\mid {E_{i,f}}\mid ^{2}\ \int_{\tau _{i}}^{\tau_{f}
}d\tau ^{\prime }\int_{\tau _{i}}^{\tau_{f} }d\tau ^{\prime \prime}e^{
-i(E_{f}-E_{i})(\tau ^{\prime }-\tau ^{\prime \prime })}iG(\tau ^{\prime
\prime },\tau ^{\prime })\hspace{0.15cm}  \label{v}
\end{equation}%
where $iG(\tau ^{\prime \prime },\tau ^{\prime })=\left\langle i\right\vert
\varphi (x(\tau ^{\prime \prime }))\varphi (x(\tau ^{\prime }))\left\vert
i\right\rangle $ is the Wightman function. For the special case in which $%
\left\vert i\right\rangle $ corresponds to a Lorentz invariant vacuum, the
transition amplitude ${\cal M}(i,f)$ vanishes for a detector moving with a
constant velocity so the detection of particles is due to acceleration. Now
we want to examine the case $\left\vert i\right\rangle =\left\vert
0\right\rangle $ where $\left\vert 0\right\rangle $ is the preferred vacuum
introduced in the previous section. In (\ref{u}), $\left\langle f\right\vert
\varphi (x)\left\vert i\right\rangle \hspace{0.15cm}$ vanishes at the first order except for $%
\left\vert f\right\rangle =\left\vert 1_{k}\right\rangle $ so
\begin{equation}
\left\langle 1_{k}\right\vert \varphi (x)\left\vert 0\right\rangle =\frac{1}{%
(2\pi )^{3/2}}\int_{|\vec{k'}|\leq k_c} d^{3}k^{\prime
}\frac{1}{\sqrt{2\omega }}\left\langle
1_{k}\right\vert a_{k^{\prime }}^{\dag }\left\vert 0\right\rangle e^{-i\vec{%
k^{\prime }}.\vec{x^{\prime }}}e^{i{\omega }^{\prime }t}=\frac{1}{(2\pi
)^{3/2}}e^{-i\vec{k}.\vec{x}}e^{i{\omega }t}\hspace{0.15cm}  \label{w}
\end{equation}%
where $\omega ^{2}=k^{2}-\alpha ^{2}k^{4}+m^2$. For a detector with
constant velocity $\vec{v}$ with respect to the preferred frame, the
world line is
\begin{equation}
\vec{x}=\vec{x}_{0}+\vec{v}t=\vec{x}_{0}+\vec{v}\tau (1-v^{2})^{-1/2}\hspace{%
0.15cm}  \label{x}
\end{equation}%
where $\vec{x}_{0}$ and $\vec{v}$ are constants and${\cal \ }\left\vert \vec{%
v}\right\vert <1{\cal \ }$($c=1$). In this case the transition amplitude $%
{\cal M}(i,f)$ takes the form
\begin{equation}
\begin{array}{c}
\vspace{0.15cm}{\cal M}(i,f)=\frac{1}{\sqrt{4\pi \omega }}\beta
E_{i,f}\int_{-\infty }^{+\infty }d\tau \hspace{0.15cm}e^{-i(E_{f}-E_{0})\tau
}\hspace{0.15cm}e^{i\tau (\omega -\vec{k}.\vec{v})(1-v^{2})^{-1/2}} \\
\vspace{0.15cm}\hspace{0.7cm}=\frac{1}{\sqrt{4\pi \omega }}\beta E_{i,f}%
\hspace{0.15cm}\delta (E_{f}-E_{0}+(\omega -\vec{k}.\vec{v})(1-v^{2})^{-1/2})
\hspace{0.15cm}.%
\end{array}
\label{y}
\end{equation}%
In the causal sector satisfying the energy positivity condition this
transition amplitude vanishes for any $E_f>E_0$ because in this
sector all wave-vectors are time-like which means $\omega$ is always
greater than $|\vec{k}.\vec{v}|$. Therefore no particle can be
detected.

In the massless limit the situation is different. In this case the
cutoff $k_c$ approaches to zero. Thus with this choice of the cutoff
there is no causal sector. But one can take a new cutoff to study
the massless limit, namely $k_c=\sqrt{3}/(2\alpha)$. This cutoff
eliminates all super-luminal group velocities and all imaginary
modes (For the massive case this cutoff does not ensures the energy
positivity condition). However one can not satisfy the energy
positivity because all wave-vectors are space-like. Thus in the
massless limit we deal with a theory not exhibiting energy
positivity. This violation of energy positivity leads to a
non-vanishing transition amplitude for the detector. In fact, by
choosing an appropriate $\vec{v}$ the frequency $\omega$ becomes
smaller than $|\vec{k}.\vec{v}|$. It implies a non-vanishing
transition amplitude for some $E_f>E_0$ in (\ref{y}), provided $k$
satisfies the following inequality
\begin{equation}
k^2>\frac{1-v^{2}\cos ^{2}\theta+}{\alpha^2 }\hspace{0.15cm}
\label{z}
\end{equation}%
where $\theta $ is the angle between $\vec{k}$ and
$\vec{v}$.\newline It means that a detector moving with a constant
velocity with respect to the preferred frame has a non-vanishing
response, i.e. it registers quanta of the field $\varphi$.

At this point the question arises: what is the physical
interpretation of this non-vanishing response. The violation of
energy positivity in the massless limit of the model leads to a
stability problem because a space-like energy momentum four vector
with a negative 0th component can be converted, under a suitable
boost transformation, to a one with a positive 0th component. Thus
we can not have a lower bound for the energy of the system in all
inertial frames. It means that the vacuum state in this case is
instable. This is the origin of the non-vanishing response of the
detector.
\section{Conclusions}

As we have seen, the response of a detector computed in the
preferred vacuum of the Lorentz symmetry breaking model is
non-vanishing when the causal sector violates the energy positivity
condition. But, it must be noticed that the detection of particles
is very sensible to the form of the modified dispersion relation and
the causal sector. For instance by means of a non-vanishing mass
parameter one can construct a causal sector which satisfies the
energy positivity condition. As it has been shown, in this case no
particle creation occurs.

\vspace{5mm}\noindent\\
{\bf Acknowledgements}\vspace{1mm}\noindent\\
The authors would like to thank H.R. Sepangi for his careful reading
of the manuscript. N. Khosravi and E. Khajeh thank H. Salehi for
encouragements. N. Khosravi also thanks the research office of
Shahid Beheshti University for financial supports.

\end{document}